\documentclass[5p,twocolumn,times]{elsarticle}
\biboptions{sort&compress}
\journal{Physics Letters B}

\newcommand{\ifproofpre}[2]{#1}

\usepackage{graphicx}
\usepackage{amsmath}
\usepackage{amssymb}
\usepackage{mathtools}
\usepackage{isotope}
\usepackage{dcolumn}  
\newcolumntype{x}[1]{D{.}{.}{#1}}
\usepackage{mathrsfs}

\usepackage{liealg}
\usepackage{wignert}
\usepackage{mcmath}



\newcommand{\Nmax}{{N_\text{max}}}
\newcommand{\scrD}{{\mathscr{D}}}

\newcommand{\scrR}{{\mathscr{R}}}
\newcommand{\MeV}{{\mathrm{MeV}}}

\begin{document}


\title{Emergence of rotational bands in \textit{ab initio} no-core \\ configuration interaction calculations of light nuclei}

\author[nd]{M. A. Caprio}
\author[iastate]{P. Maris}
\author[iastate]{J. P. Vary}
\address[nd]{Department of Physics, University of Notre Dame, Notre Dame, Indiana 46556-5670, USA}
\address[iastate]{Department of Physics and Astronomy, Iowa State University, Ames, Iowa 50011-3160, USA}

\date{\today} 

\begin{abstract}
The emergence of rotational bands is observed in no-core configuration
interaction (NCCI) calculations for the odd-mass $\isotope{Be}$ isotopes
($7\leq A \leq 13$) with the JISP16 nucleon-nucleon interaction, as evidenced by rotational patterns for
excitation energies, quadrupole moments, and $E2$ transitions.  Yrast and
low-lying excited bands are found.  The results demonstrate the
possibility of well-developed rotational structure in NCCI
calculations using a realistic nucleon-nucleon interaction.
\end{abstract}



\maketitle

\section{Introduction}
\label{sec-intro}

Nuclei exhibit a wealth of collective phenomena, including clustering,
rotation, and
pairing~\cite{rowe2010:collective-motion,bohr1998:v1,bohr1998:v2}.
Collective dynamics have been extensively modeled in phenomenological
descriptions~\cite{rowe2010:collective-motion,bohr1998:v2,iachello1987:ibm,eisenberg1987:v1}.
Some forms of collectivity may also be obtained microscopically in the
conventional (valence) shell model, \textit{e.g.}, Elliott $\grpsu{3}$
rotation~\cite{elliott1958:su3-part1,harvey1968:su3-shell}.  However,
observing the emergence of collective phenomena directly from first
principles~--- that is, in a fully \textit{ab initio} calculation of
the nucleus, as a many-body system in which all the constituent
protons and neutrons participate, with realistic
interactions~--- remains as an outstanding challenge.

Recent developments in large-scale calculations have brought
significant progress in the $\textit{ab initio}$ description of light
nuclei~\cite{pieper2004:gfmc-a6-8,neff2004:cluster-fmd,hagen2007:coupled-cluster-benchmark,navratil2009:ncsm,bacca2012:6he-hyperspherical}.
In \textit{ab initio} no-core configuration interaction (NCCI)
approaches~--- such as the no-core shell model
(NCSM)~\cite{navratil2000:12c-ab-initio,navratil2000:12c-ncsm,navratil2009:ncsm,vary2009:ncsm-mfdn-scidac09,barrett:ncsm}, no-core Monte Carlo shell model (MCSM)~\cite{abe2012:fci-mcsm-ncfc},
or no-core full configuration (NCFC)~\cite{maris2009:ncfc} methods~---
the nuclear many-body bound-state eigenproblem is formulated as a
matrix diagonalization problem.  The Hamiltonian is represented with
respect to a basis of antisymmetrized products of single-particle
states, generally harmonic oscillator states, and the problem is
solved for the full system of $A$ nucleons, \textit{i.e.}, with no
inert core.  In practice, such calculations must be carried out in a
finite space, obtained by truncating the many-body basis according to
a maximum allowed number $\Nmax$ of oscillator excitations above the
lowest oscillator configuration (\textit{e.g.},
Ref.~\cite{navratil2009:ncsm}).  With increasing $\Nmax$, the results
converge towards those which would be achieved in the full,
infinite-dimensional space for the many-body system.

Computational restrictions limit the extent to which converged
calculations can be obtained for the observables needed for the
identification of collective phenomena.  In particular, the observables
most indicative of rotational collectivity~--- $E2$ matrix
elements~--- present special challenges for convergence in an NCCI
approach~\cite{bogner2008:ncsm-converg-2N,cockrell2012:li-ncfc}, due
to their sensitivity to the large-radius asymptotic portions of the
nuclear wave function.  Nonetheless, some promising suggestions of
collective phenomena, \textit{e.g.}, deformation and
clustering, have already been obtained in \textit{ab initio}
calculations~\cite{wiringa2000:gfmc-a8,neff2008:clustering-nuclei,cockrell2012:li-ncfc,kanadaenyo2012:amd-cluster,shimizu2012:mcsm,maris2012:mfdn-ccp11}.

In this letter, we observe the emergence of collective rotation in
\textit{ab initio} 
NCCI calculations for the $\isotope{Be}$ isotopes, using the realistic
JISP16 nucleon-nucleon interaction~\cite{shirokov2007:nn-jisp16}.
Evidence for rotational band structure is found in the calculated
excitation energies, quadrupole moments, and $E2$ transition matrix
elements.  In NCCI calculations of the even-mass $\isotope{Be}$
nuclei, yrast sequences of angular momenta $0$, $2$, $4$, $\ldots$
arise with calculated properties resembling those of $K=0$
ground-state rotational bands (see Ref.~\cite{maris2012:mfdn-hites12}
for a preliminary report of comparable results for $\isotope[12]{C}$).
However, the most distinctive, well-developed, and systematic
rotational band structures are observed in calculations for
odd-mass nuclei. Given the same range of excitation energies and
angular momenta, the low-lying $\Delta J=1$ bands in the odd-mass
nuclei provide a richer set of energy and electromagnetic observables.
We therefore focus here on the odd-mass $\isotope{Be}$ isotopes,
specifically, with $7\leq A
\leq 13$.  After a brief review of the properties expected in nuclear rotational structure
(Sec.~\ref{sec-rot}), the results for rotational bands in NCCI
calculations of these $\isotope{Be}$ isotopes are presented (Sec.~\ref{sec-results}).  
Preliminary results for $\isotope[9]{Be}$ were reported in Ref.~\cite{maris2012:mfdn-ccp11}.

\section{Rotation}
\label{sec-rot}

We first review nuclear collective
rotation and its expected signatures~\cite{rowe2010:collective-motion,bohr1998:v2}.
Under the assumption of
adiabatic separation of the rotational degree of freedom,  a nuclear state may be described in terms of an 
\textit{intrinsic state}, as viewed in the non-inertial intrinsic frame, together with the rotational
motion of this intrinsic frame.    For axially symmetric structure, in particular, the
intrinsic state $\tket{\phi_K}$ is characterized by definite angular momentum projection $K$ along
the intrinsic symmetry axis.  The full nuclear state $\tket{\psi_{JKM}}$,
with total angular momentum $J$ and projection $M$, then has the form
\begin{multline}
\label{eqn-psi}
\tket{\psi_{JKM}}=\Bigl[\frac{2J+1}{16\pi^2(1+\delta_{K0})}\Bigr]^{1/2}
\int d\vartheta\,\bigl[\scrD^J_{MK}(\vartheta)\tket{\phi_K;\vartheta}
\ifproofpre{\\}{}
+(-)^{J+K}\scrD^J_{M-K}(\vartheta)\tket{\phi_{\bar{K}};\vartheta}\bigr],
\end{multline}
where $\vartheta$ represents the Euler angles for rotation of the
intrinsic state, and $\tket{\phi_{\bar{K}}}$ is the $\scrR_2$-conjugate
intrinsic state, which has angular momentum projection $-K$ along the
symmetry axis.  

The most recognizable features in the spectroscopy of
rotational states reside not in the states taken individually but in the relationships among the different
states $\tket{\psi_{JKM}}$ sharing the same intrinsic state
$\tket{\phi_K}$.  These
states  constitute members of a rotational band, with angular
momenta $J=K$, $K+1$, $\ldots$, except with only
even $J$ (or only odd $J$, depending upon the intrinsic $\scrR_2$ symmetry) for $K=0$ bands.
Within a rotational band, energies follow the pattern
\begin{equation}
\label{eqn-EJ}
E(J)=E_0+AJ(J+1),
\end{equation}
where, in terms of the moment of inertia $\cal{J}$
about an axis perpendicular to the symmetry axis, $A\equiv\hbar^2/(2\cal{J})$.  For $K=1/2$ bands,
the Coriolis contribution to the kinetic energy results in an
energy staggering given by
\begin{equation}
\label{eqn-EJ-stagger}
E(J)=E_0+A\bigl[J(J+1)+a(-)^{J+1/2}(J+\tfrac12)\bigr],
\end{equation}
where $a$ is the Coriolis decoupling parameter.  Reduced matrix
elements $\trme{\psi_{J_fK}}{Q_2}{\psi_{J_iK}}$ of the electric
quadrupole operator $Q_2$ between states within a band are entirely
determined by the rotational structure, except for the overall
normalization, which is proportional to the intrinsic quadrupole
moment $eQ_0\equiv(16\pi/5)^{1/2}\tme{\phi_K}{Q_{2,0}}{\phi_K}$.  In
particular, quadrupole moments within a band are obtained as
\begin{equation}
\label{eqn-Q}
Q(J)=\frac{3K^2-J(J+1)}{(J+1)(2J+3)}Q_0,
\end{equation}
and reduced transition probabilities as
\begin{equation}
\label{eqn-BE2}
B(E2;J_i\rightarrow J_f)=\frac{5}{16\pi} \tcg{J_i}{K}{2}{0}{J_f}{K}^2 (eQ_0)^2.
\end{equation}
In obtaining these results, $Q_2$ can be taken to be any operator of the form $Q_{2\mu}=\sum_{i=1}^Ae_i
r_{i}^2Y_{2\mu}(\uvec{r}_{i})$ and may therefore represent the electromagnetic $E2$
transition operator, mass quadrupole tensor, neutron quadrupole
tensor, \textit{etc.}, depending upon the choice of coefficients $e_i$
(see Sec.~\ref{sec-results}).

\section{Results}
\label{sec-results}

An NCCI calculation is defined by the interaction for the nuclear system and by the truncated many-body
space in which the calculation is carried out.  The present 
calculations use the JISP16
interaction~\cite{shirokov2007:nn-jisp16}, which is a two-body
interaction derived from neutron-proton scattering data and adjusted
via a phase-shift equivalent transformation to describe light nuclei
without explicit three-body interactions.  The bare interaction
is used, without renormalization to the truncated
space~\cite{maris2009:ncfc}.  The Coulomb interaction has been omitted
from the Hamiltonian, to ensure exact conservation of isospin, thereby
simplifying the spectrum.  (The primary effect of the Coulomb
interaction, when included, is to induce a shift in the overall
binding energies, which is irrelevant to identification of rotational band structure.)
These calculations are carried out for oscillator
truncations ranging from $\Nmax=10$ for $\isotope[7]{Be}$ to $\Nmax=7$
for $\isotope[13]{Be}$, with basis oscillator 
$\hbar\Omega$ parameters near the variational minimum
($\hbar\Omega=20\text{--}22.5\,\MeV$).  The proton-neutron $M$-scheme
code
MFDn~\cite{sternberg2008:ncsm-mfdn-sc08,maris2010:ncsm-mfdn-iccs10,aktulga2012:mfdn-ep2012}
has been used for the many-body calculations.

\begin{figure}[t]
\begin{center}
\includegraphics*[width=\ifproofpre{0.90}{0.65}\hsize]{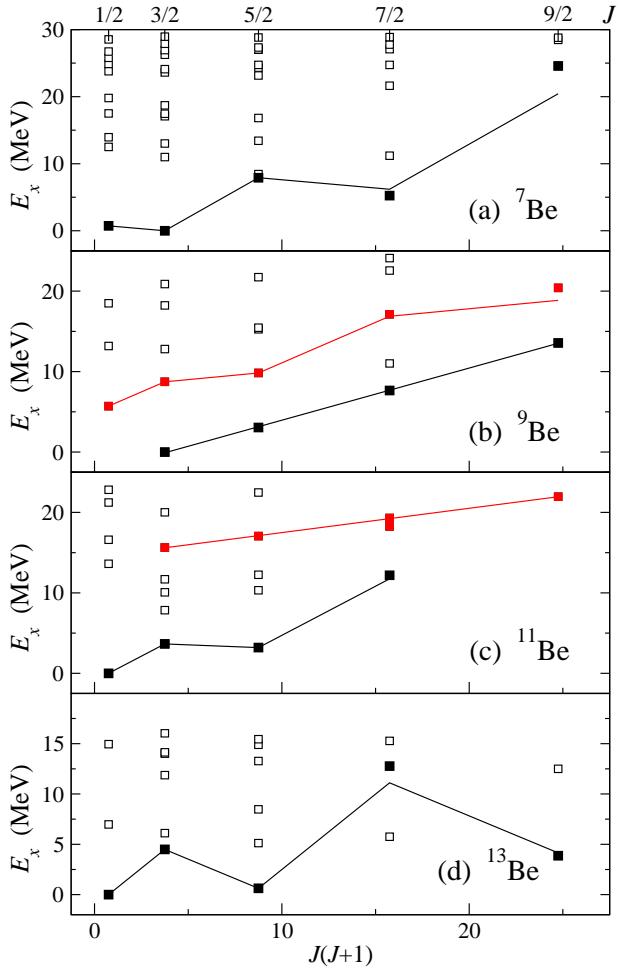}
\end{center}
\caption{Excitation energies obtained for states in  the
\textit{natural} parity spaces of the odd-mass $\isotope{Be}$ isotopes: (a)~$\isotope[7]{Be}$, (b)~$\isotope[9]{Be}$,
(c)~$\isotope[11]{Be}$, and (d)~$\isotope[13]{Be}$.  Energies are
plotted with respect to $J(J+1)$ to facilitate identification of
rotational energy patterns, while the $J$ values themselves are
indicated at top.  Filled symbols indicate candidate rotational
bandmembers (black for yrast states and red for excited states).  The
lines indicate the corresponding best fits for rotational energies.
Where quadrupole transition strengths indicate significant two-state
mixing (see text), more than one state of a given $J$ is indicated as a
bandmember.
}
\label{fig-Ex-odd-nat}
\end{figure}

\begin{figure}[t]
\begin{center}
\includegraphics*[width=\ifproofpre{0.90}{0.65}\hsize]{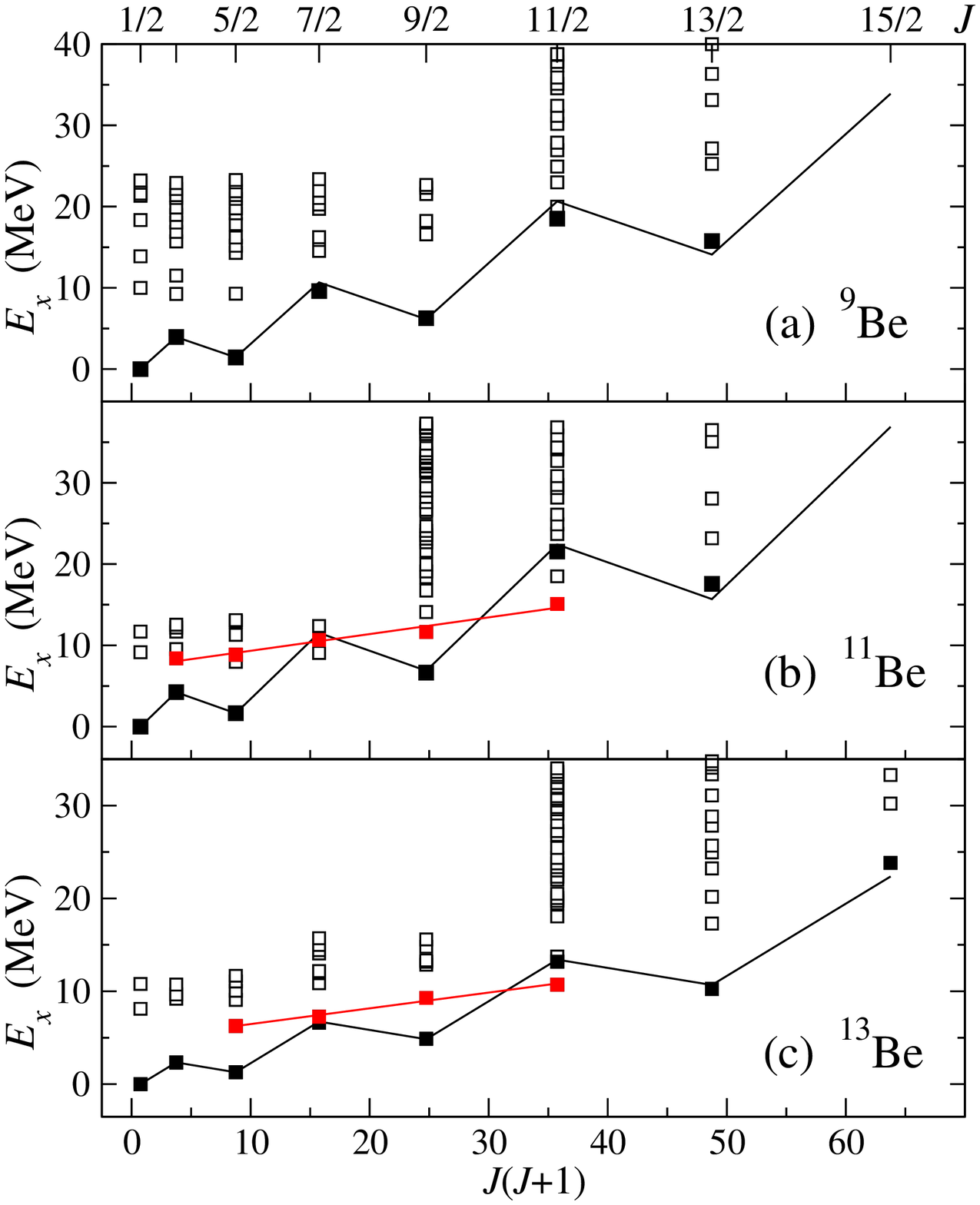}
\end{center}
\caption{Excitation energies obtained for states in  the
\textit{unnatural} parity spaces of the odd-mass $\isotope{Be}$ isotopes: (a)~$\isotope[9]{Be}$,
(b)~$\isotope[11]{Be}$, and (c)~$\isotope[13]{Be}$.  
Energies are
plotted with respect to $J(J+1)$ to facilitate identification of
rotational energy patterns, while the $J$ values themselves are
indicated at top.  Filled symbols indicate candidate rotational
bandmembers (black for yrast states and red for excited states).  The
lines indicate the corresponding best fits for rotational energies.
Quadrupole transition strengths indicate significant but ambiguous two-state or multistate
mixing for certain poorly-isolated
bandmembers.}
\label{fig-Ex-odd-un}
\end{figure}

The calculated excitation energies for low-lying states of the
odd-mass $\isotope{Be}$ isotopes (with $7\leq A \leq 13$) are shown in
Figs.~\ref{fig-Ex-odd-nat} and~\ref{fig-Ex-odd-un}. For each nucleus,
there are two parity spaces to consider, shown separately in these two
figures (energies are taken relative to the lowest state of the same
parity).  We refer to the parity of the lowest allowed oscillator
configuration (negative for $\isotope[7,9,11]{Be}$, positive for
$\isotope[13]{Be}$) as the
\textit{natural parity} (Fig.~\ref{fig-Ex-odd-nat}) and that obtained
by promoting one nucleon by one shell as the \textit{unnatural parity}
(Fig.~\ref{fig-Ex-odd-un}).  The NCCI bases for these spaces consist
of states with even and odd numbers of oscillator excitations,
respectively, above the lowest configuration.  While the lowest
unnatural parity states normally lie at significantly higher energy
than those of natural parity,
they are calculated
to lie within a few $\MeV$ of the lowest natural parity states in the
isotopes $\isotope[9,11,13]{Be}$~\cite{maris2012:mfdn-ccp11}, which are therefore included in
Fig.~\ref{fig-Ex-odd-un}.  Note that parity inversion arises for
$\isotope[11]{Be}$, \textit{i.e.}, the ground state is experimentally~\cite{npa1990:011-012}
in the unnatural parity space, and both spaces are near-degenerate in
calculations at finite $\Nmax$ (see Ref.~\cite{navratil2009:ncsm}).
The minimal isospin ($T=T_z$) spectrum is shown in each case.

\begin{figure}[t]
\begin{center}
\includegraphics*[width=\ifproofpre{0.90}{0.65}\hsize]{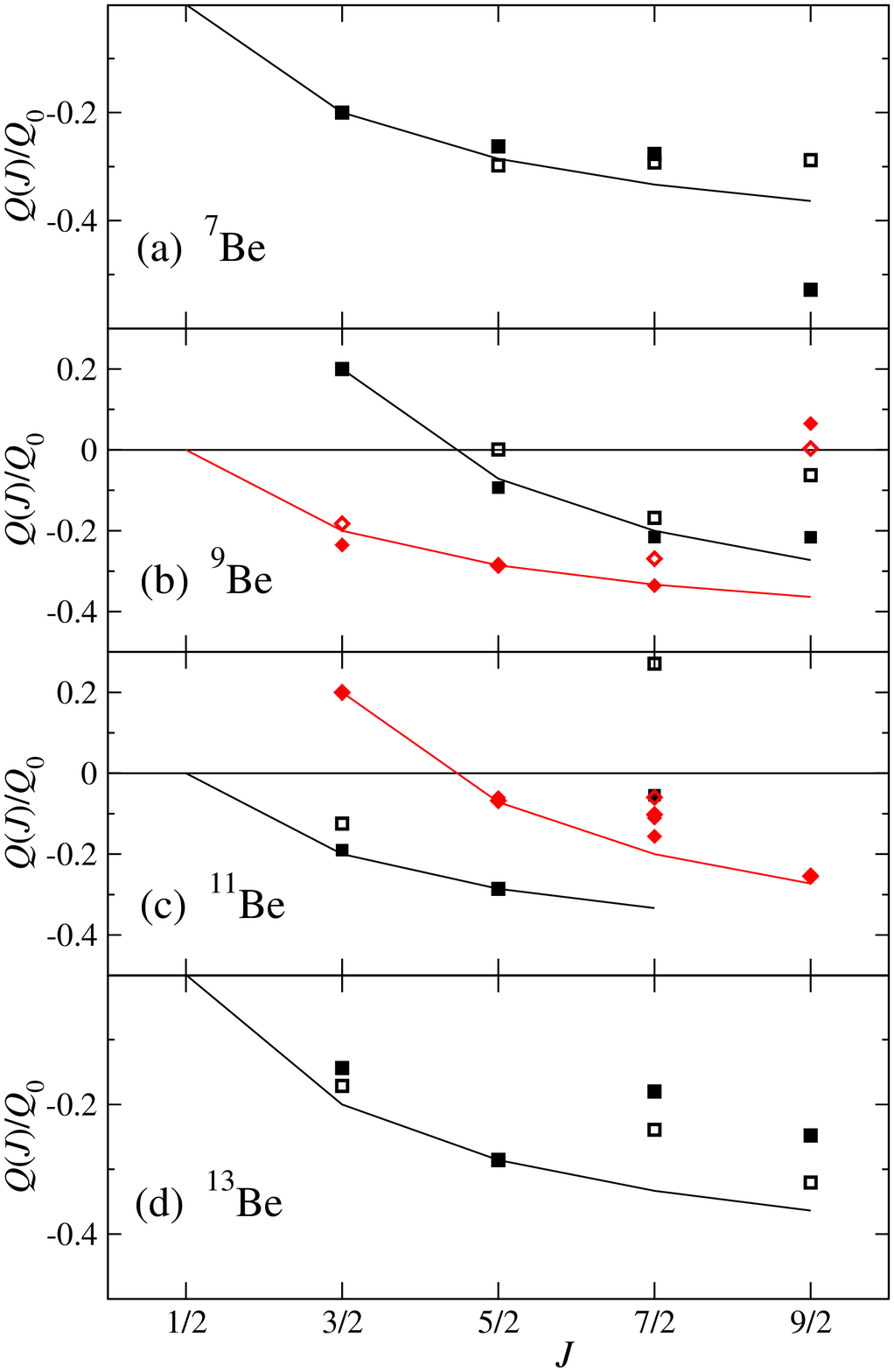}
\end{center}
\caption{Quadrupole moments calculated for candidate bandmembers in the
\textit{natural} parity spaces of the odd-mass $\isotope{Be}$ isotopes: (a)~$\isotope[7]{Be}$, (b)~$\isotope[9]{Be}$,
(c)~$\isotope[11]{Be}$, and (d)~$\isotope[13]{Be}$.   The states are as
identified in Fig.~\ref{fig-Ex-odd-nat} and are shown as
black squares for yrast states or red diamonds for excited states.
Filled symbols indicate proton quadrupole moments, and open symbols
indicate neutron quadrupole moments.  The curves indicate the theoretical values for a
$K=1/2$ or $K=3/2$ rotational band, as appropriate, given by~(\ref{eqn-Q}).  Quadrupole moments are
normalized to $Q_0$, which is defined by either the $J=3/2$ or $J=5/2$
bandmember (see text). 
}
\label{fig-Q-odd-nat}
\end{figure}
\begin{figure}[t]
\begin{center}
\includegraphics*[width=\ifproofpre{0.90}{0.65}\hsize]{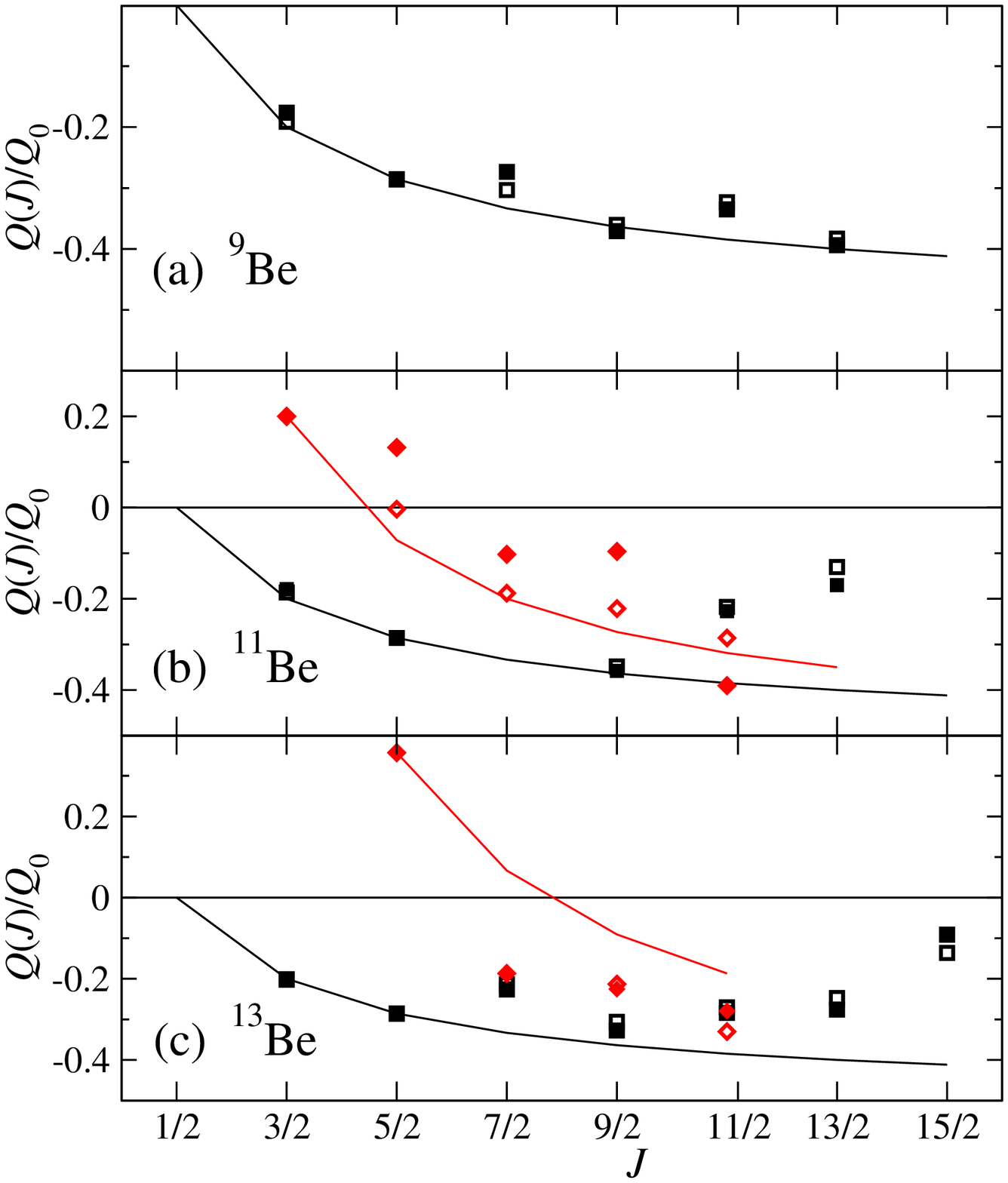}
\end{center}
\caption{Quadrupole moments calculated for candidate bandmembers in the
\textit{unnatural} parity spaces of the odd-mass $\isotope{Be}$ isotopes: (a)~$\isotope[9]{Be}$,
(b)~$\isotope[11]{Be}$, and (c)~$\isotope[13]{Be}$.
The states are as
identified in Fig.~\ref{fig-Ex-odd-un} and are shown as
black squares for yrast states or red diamonds for excited states. Filled symbols indicate proton quadrupole moments, and open symbols
indicate neutron quadrupole moments.  The curves indicate the theoretical values for a
$K=1/2$ or $K=3/2$ rotational band, as appropriate, given by~(\ref{eqn-Q}). Quadrupole moments are
normalized to $Q_0$, which is defined by either the $J=3/2$ or $J=5/2$
bandmember (see text). 
}
\label{fig-Q-odd-un}
\end{figure}

\begin{figure}[t]
\begin{center}
\includegraphics*[width=\ifproofpre{0.90}{0.65}\hsize]{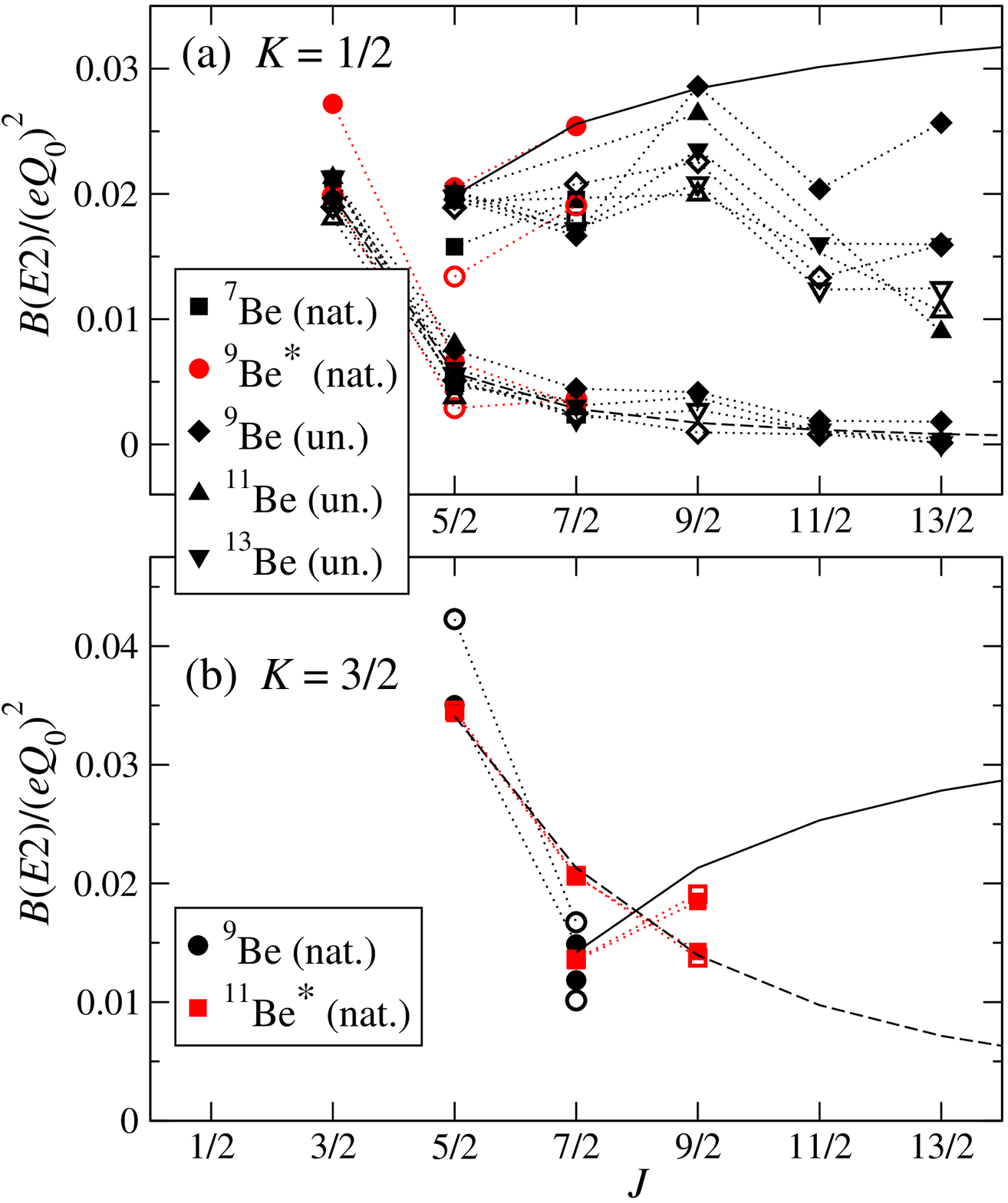}
\end{center}
\caption{Transition $B(E2;J\rightarrow J-2)$ and $B(E2;J\rightarrow
J-1)$ strengths calculated between candidate bandmembers, for bands
with (a)~$K=1/2$ or (b)~$K=3/2$, in the odd-mass $\isotope{Be}$
isotopes $\isotope[7]{Be}$, $\isotope[9]{Be}$, $\isotope[11]{Be}$, and
$\isotope[13]{Be}$ (asterisks in the legend indicate excited bands).
The curves indicate the theoretical values for $\Delta J=2$ (upper
curves, solid)
and $\Delta J=1$ (lower curves, dashed) transitions in a band of the given $K$,
given by~(\ref{eqn-BE2}).  Calculated transition strengths of the same
$\Delta J$ within a band are connected by dotted lines.  Filled
symbols indicate proton strengths, and open symbols indicate neutron
strengths.  Values are normalized to the same intrinsic quadrupole
moment $Q_0$ as in Fig.~\ref{fig-Q-odd-nat}
or~\ref{fig-Q-odd-un}. Where transition strengths indicate significant
fragmentation of the rotational state over multiple calculated states,
namely, for the $\isotope[11]{Be}$ excited $J=7/2$ state identified in
Fig.~\ref{fig-Ex-odd-nat}(c), the $B(E2)$ values for transitions into
or out of this state are summed.  }
\label{fig-BE2-odd}
\end{figure}

To facilitate identification of rotational bands, it is helpful to
plot the calculated excitation energies with respect to $J(J+1)$, so
that energies within an ideal rotational band would lie on a straight
line~--- or, for $K=1/2$ bands, staggered about a straight line.  For
the candidate $K=1/2$ bands in Figs.~\ref{fig-Ex-odd-nat}
and~\ref{fig-Ex-odd-un}, an energy fit is obtained by adjusting the
parameters of~(\ref{eqn-EJ-stagger}) to the first three bandmembers
(the remainder of the line is thus an extrapolation). For the
remaining bands, a straight line fit of~(\ref{eqn-EJ}) to all
bandmembers is shown.

The yrast and near-yrast states yield the most immediately
recognizable sets of candidate bandmembers.  Yrast rotational bands
(with bandmembers indicated by solid black squares in
Figs.~\ref{fig-Ex-odd-nat} and~\ref{fig-Ex-odd-un}) are found
with $K=1/2$ except in the natural parity space of
$\isotope[9]{Be}$ [Fig.~\ref{fig-Ex-odd-nat}(b)], for which the yrast
band has $K=3/2$.  The density of states rapidly increases off the
yrast line, hindering identification of candidate bands and
furthermore suggesting that the rotational states may be fragmented by
mixing with nearby states.  Nonetheless, several excited candidate
bands (indicated by solid red squares) can also be clearly identified,
with $1/2\leq K \leq 5/2$, once transition strengths have been taken
into account.

For the yrast $K=1/2$ bands, as a result of Coriolis decoupling,
it should be noted that alternate bandmembers are raised in energy into a region of higher
density of states, which complicates identification and is conducive
to fragmentation.  The energy staggering in the calculated yrast band of
the $\isotope[7]{Be}$ natural parity space
[Fig.~\ref{fig-Ex-odd-nat}(a)]~--- in which the $J=3/2$, $7/2$, $\ldots$ levels are lowered, and the
$J=1/2$, $5/2$, $\ldots$ levels are raised~--- corresponds to a negative value of the decoupling
parameter.  Note that the staggering is
sufficiently pronounced that the two lowest-$J$ bandmembers are
inverted, as experimentally observed for this
nucleus~\cite{npa2002:005-007}.  Then, positive values
of the decoupling parameter are instead obtained for the
remaining $K=1/2$ bands.

It is interesting to compare these \textit{ab initio} results for the
yrast bands in the natural parity spaces (Fig.~\ref{fig-Ex-odd-nat})
with the Nilsson model
predictions~\cite{nilsson1955:model,mottelson1959:nilsson-model}.
Specifically, the
calculated yrast bands have $K=1/2$ ($a\approx-1.4$) for $\isotope[7]{Be}$,
$K=3/2$ for $\isotope[9]{Be}$, $K=1/2$ ($a\approx+1.2$) for
$\isotope[11]{Be}$, and $K=1/2$ ($a\approx+3.1$) for
$\isotope[13]{Be}$.  The expected
Nilsson $[Nn_z\Lambda\Omega]$ asymptotic quantum number assignments
(see Fig.~5-1 of
Ref.~\cite{bohr1998:v2}) and corresponding Nilsson values of the
decoupling parameter are $[110\tfrac12]$ ($a\approx-1$) for
$\isotope[7]{Be}$, $[101\tfrac32]$ for $\isotope[9]{Be}$,
$[101\tfrac12]$ ($a\approx0$) for $\isotope[11]{Be}$, and
$[220\tfrac12]$ ($a\approx+1$) for $\isotope[13]{Be}$.
We see consistency not only in the $K$ ($=\Omega$) quantum numbers for the band but also in the
\textit{qualitative} trend of the decoupling parameters for these bands.
(The Nilsson
values for $a$~\cite{nilsson1955:model,mottelson1959:nilsson-model} consider
mixing of spherical orbitals only within a single spherical oscillator
shell, which is sufficient for a weakly-deformed oscillator-like mean
field.  However, they should not be expected to provide \textit{quantitative}
accuracy for a nucleon in, say, the mean field produced by a double-$\alpha$
$\isotope[8]{Be}$ core.)

The quadrupole moments for all states within the candidate bands are
shown in Figs.~\ref{fig-Q-odd-nat} and~\ref{fig-Q-odd-un}, both for
the yrast bands (black squares) and for the excited bands (red
diamonds).  The values are normalized to $Q_0$, to facilitate
comparison with the rotational predictions for $Q(J)/Q_0$
from~(\ref{eqn-Q}) (shown as curves in each plot).  The value of $Q_0$ used
for normalization has in each case been obtained simply from the
quadrupole moment of the lowest-energy bandmember of nonvanishing
quadrupole moment.  (Thus, for $K=1/2$ bands, since the
quadrupole moment of the $J=1/2$ bandhead vanishes identically, either the $J=3/2$ or $5/2$ bandmember is used for
normalization, according to the staggering.)
Quadrupole moments in Figs.~\ref{fig-Q-odd-nat} and~\ref{fig-Q-odd-un}
are calculated using both the proton (filled
symbols) and neutron (open symbols) quadrupole tensors.\footnote{The proton
quadrupole tensor, defined as
$Q_{2\mu,p}=\sum_{i=1}^Zr_{p,i}^2Y_{2\mu}(\uvec{r}_{p,i})$, is the
operator used in calculation of the physically observable
electromagnetic moments and transitions.  However, the rotational
relations~(\ref{eqn-Q}) and~(\ref{eqn-BE2}) are equally
applicable to matrix elements of the neutron quadrupole tensor,
$Q_{2\mu,n}=\sum_{i=1}^Nr_{n,i}^2Y_{2\mu}(\uvec{r}_{n,i})$.
These therefore provide a valuable complementary
set of observables for purposes of investigating whether or
not the nuclear wave functions satisfy the conditions of adiabatic
rotational separation, particularly relevant, due to the high
neutron-proton asymmetry, in the
neutron-rich $\isotope{Be}$ isotopes.}

Finally, in-band transition strengths are shown in
Fig.~\ref{fig-BE2-odd}, again as obtained for both proton (solid
symbols) and neutron (open symbols) quadrupole operators, and for
$\Delta J=2$ transitions (upper curves, solid) and $\Delta J=1$
transitions (lower curves, dashed).  The various $K=1/2$ bands are
superposed in Fig.~\ref{fig-BE2-odd}(a), and the $K=3/2$ bands are
shown in Fig.~\ref{fig-BE2-odd}(b).  Transition strengths are
normalized as $B(E2;J\rightarrow J-\Delta J)/(eQ_0)^2$, for comparison
with the rotational values from~(\ref{eqn-BE2}).  The same $Q_0$
values are used as in Figs.~\ref{fig-Q-odd-nat}
and~\ref{fig-Q-odd-un}, \textit{i.e.}, obtained from $Q(3/2)$ or $Q(5/2)$.
Therefore, no free normalization parameters remain for the $B(E2)$
strengths in Fig.~\ref{fig-BE2-odd}.  For instance, it may be observed
that the values for $B(E2;3/2\rightarrow 1/2)/(eQ_0)^2$ in
Fig.~\ref{fig-BE2-odd}(a) cluster at the rotational value, indicating that the calculated $B(E2;3/2\rightarrow 1/2)$ strengths are
in the proper relation to the calculated
$Q(3/2)$ moment or $Q(5/2)$ moment, as appropriate,
consistent with adiabatic rotation.

The level of resemblance between the calculated energies, quadrupole
moments, and transition strengths for the candidate bands and
the expected rotational values in
Figs.~\ref{fig-Ex-odd-nat}--\ref{fig-BE2-odd}, while clearly not
perfect, indicates a remarkably clean separation
of rotational and intrinsic degrees of freedom in these \textit{ab
initio} NCCI
calculations.  One should bear in mind that quadrupole moments of
\textit{arbitrarily} chosen states in the spectrum fluctuate not only in
magnitude but also in sign, and that calculated $E2$ strengths among arbitrarily
chosen pairs of states fluctuate by many orders of magnitude.  (The
$3/2\rightarrow 1/2$ transitions in Fig.~\ref{fig-BE2-odd} are
enhanced by factors of $\sim1.1$--$17$ relative to the typical Weisskopf
single-particle estimate~\cite{weisskopf1951:estimate}.)   

It is worth highlighting a few notable features from the band
structures in Figs.~\ref{fig-Ex-odd-nat}--\ref{fig-BE2-odd}:

(1) The $K=1/2$ yrast bands in the unnatural parity spaces
(Fig.~\ref{fig-Ex-odd-un}) can be traced to $J$ values as high as $\sim13/2$.
For instance, for
$\isotope[13]{Be}$ [Fig.~\ref{fig-Ex-odd-un}(c)], the
energies of the $J=7/2$, $9/2$, $11/2$, and $13/2$ bandmembers all
agree with the rotational values, from~(\ref{eqn-EJ-stagger}), to within
$0.4\,\MeV$, and a $J=15/2$ bandmember can also be reasonably
identified (within $1.5\,\MeV$ of the
rotational energy).  The quadrupole moments
[Fig.~\ref{fig-Q-odd-un}(c)] of the $J=3/2$ and $J=5/2$ bandmembers
(the latter is used to determine $Q_0$ in the figure) are in the
expected rotational ratio, from~(\ref{eqn-Q}), to within $1.1\%$
for protons or $0.4\%$ for neutrons.  The quadrupole moments for the
higher bandmembers are highly consistent between protons and neutrons
and have the expected sign, but they gradually fall off from the
rotational values, approaching zero for the $J=15/2$
bandmember.  

(2) For the yrast and low-lying rotational bands in the natural parity
spaces (Fig.~\ref{fig-Ex-odd-nat}), rotational behavior appears to
terminate at generally lower angular momentum.  For instance, for
$\isotope[11]{Be}$ [Fig.~\ref{fig-Ex-odd-nat}(c)], the $K=1/2$ yrast
band terminates at $J=7/2$ on the basis of energies: the lowest
calculated $J=7/2$ state lies within $0.5\,\MeV$ of the expected
energy extrapolated for an yrast bandmember, but the lowest calculated
$J=9/2$ state is $11\,\MeV$ too high in energy to be an yrast
bandmember.  The terminating angular momentum expected in a simple
valence $p$-shell or NCCI $\Nmax=0$ description is, in fact, $J=7/2$.
The quadrupole moments [Fig.~\ref{fig-Q-odd-nat}(c)] suggest that the
viability of a rotational description may end even earlier, at
$J=5/2$.  Similar comments may be made about the yrast and excited
bands in $\isotope[7]{Be}$ and $\isotope[9]{Be}$
[Fig.~\ref{fig-Ex-odd-nat}(a,b)], where the quadrupole moments
[Fig.~\ref{fig-Q-odd-nat}(a,b)] are in close agreement with rotational
values through $J=7/2$, but then begin to deviate significantly at
$J=9/2$.

(3) To some extent in the quadrupole moments, but especially in the
$\Delta J=2$ transition strengths for the $K=1/2$ bands
[Fig.~\ref{fig-BE2-odd}(a)], one may observe that the $E2$ matrix element
strengths start at the expected rotational values for low $J$
but then systematically fall off below the rotational values at higher
$J$.  This trend signals deviation from a strict \textit{adiabatic}
rotational picture, as described in Sec.~\ref{sec-rot}, but it is
also, at least qualitatively, in agreement with
more microscopic treatments of nuclear
rotation.  Specifically, $E2$ matrix elements within an Elliott
$\grpsu{3}$ band decline in strength as band termination is approached
(see
discussion in Ref.~\cite{harvey1968:su3-shell}).  A similar falloff
can be 
obtained in $\grpsptr$ symplectic
calculations~\cite{draayer1984:spsm-20ne} of rotational bands (see
Fig.~6 of Ref.~\cite{rowe1985:micro-collective-sp6r}).  Whether or not
such $\grpsu{3}$ or $\grpsptr$ rotational mechanisms are at play in
the present NCCI results awaits
full analysis in an $\grpsu{3}$/$\grpsptr$ symmetry-adapted implementation of the
NCCI approach~\cite{dytrych2008:sp-ncsm}.

As we explore the interpretation of NCCI results in a rotational
context, it is interesting to note that straightforward fragmentation
of the rotational strength over two calculated levels can be observed
for the $J=7/2$ member of the excited band in the natural parity space
of $\isotope[11]{Be}$ [Fig.~\ref{fig-Ex-odd-nat}(c)].  Transitions
into and out of this state are fragmented in the approximate
proportion $0.4:0.6$.  However, the summed strengths, shown in
Fig.~\ref{fig-BE2-odd}(b), which combine the fragmented transitions
involving this level, are in near-perfect agreement with rotational
values.

We also note that a staggering may be observed in
the $\Delta J=2$ transition strengths, to a greater or lesser degree,
for the various $K=1/2$ bands [Fig.~\ref{fig-BE2-odd}(a)].  Such
staggering is in fact consistent with the adiabatic rotational picture,
once the $\tme{\phi_K}{Q_{2,2K}}{\phi_{\bar{K}}}$ cross term in the
rotational $E2$ matrix element, neglected in~(\ref{eqn-BE2}), is taken
into account [see~(6.38) of Ref.~\cite{rowe2010:collective-motion}].
In well-deformed rotor nuclei, this contribution is commonly ignored,
on the presumption that $\tme{\phi_K}{Q_{2,0}}{\phi_K}\sim Q_0$ is
strongly enhanced while $\tme{\phi_K}{Q_{2,2K}}{\phi_{\bar{K}}}$ is of
single-particle strength~\cite{rowe2010:collective-motion}.  However,
in light nuclei, where the collective enhancement is weaker, such a
single-particle 
contribution may be expected to be nonnegligible in comparison, and to be of approximately
the magnitude seen in Fig.~\ref{fig-BE2-odd}(a).

\section{Conclusion}
\label{sec-concl}

\begin{figure}[tb]
\begin{center}
\includegraphics*[width=\ifproofpre{0.90}{0.65}\hsize]{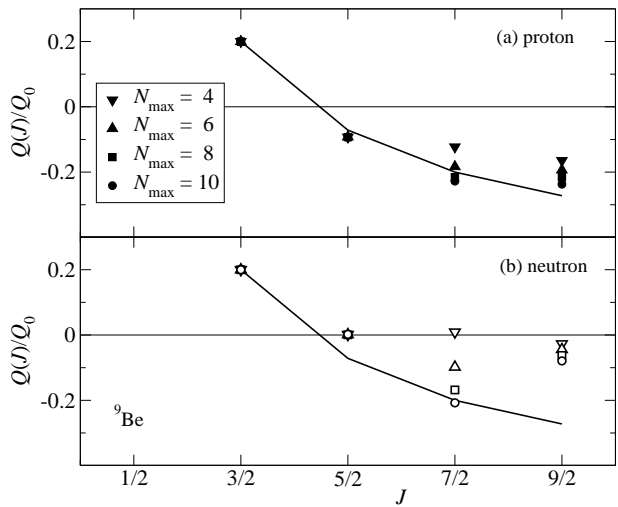}
\end{center}
\caption{Relative quadrupole moments $Q(J)/Q_0$, calculated with the (a)~proton and (b)~neutron quadrupole tensor, for the
$K=3/2$ yrast band in the $\isotope[9]{Be}$ natural parity space, from calculations with basis truncations $\Nmax=4$, $6$, $8$, and $10$.
Since the normalization $Q_0$ is obtained in each case from $Q(3/2)$,
the plotted values
indicate the $\Nmax$ dependence of the ratio $Q(J)/Q(3/2)$.
}
\label{fig-Q-Nmax-9Be}
\end{figure}

The principal challenge in identifying collective structure in NCCI
calculations with realistic interactions lies in the weak
convergence of the relevant observables.
Eigenvalues and other calculated observables are dependent
upon both the truncation $\Nmax$ and the oscillator length parameter
(or $\hbar\Omega$) for the NCCI basis.   Although it is possible
to extrapolate the values of calculated observables to their values in
the full, infinite
space~\cite{bogner2008:ncsm-converg-2N,maris2009:ncfc,cockrell2012:li-ncfc,furnstahl2012:ho-extrapolation,coon2012:nscm-ho-regulator},
such methods are still in their formative stages, especially for the
crucial $E2$ observables.
It is therefore particularly notable that 
quantitatively well-developed and robust signatures of rotation may be observed
in the present results.  That this is possible reflects the distinction
between convergence of \textit{individual} observables, taken singly, and convergence of
\textit{relative} properties, such as ratios of excitation energies or
ratios of quadrupole matrix elements.  It is these latter relative
properties which are essential to identifying rotational dynamics and
which are found to be sufficiently converged to
yield stable rotational patterns at currently achievable $\Nmax$
truncations, as illustrated in 
Fig.~\ref{fig-Q-Nmax-9Be} for the $K=3/2$ ground-state band of $\isotope[9]{Be}$.

From the results of Sec.~\ref{sec-results}, it is seen that rotational
structure is pervasive in \textit{ab initio} NCCI calculations
of light nuclei, occurring in the yrast and
near-yrast regions of all the spectra considered for the
$\isotope{Be}$ isotopic chain.  With suitable extrapolation methods in
place, a salient test of \textit{ab initio} calculations and
interactions will then be quantitative prediction of collective
rotational parameters, such as the intrinsic quadrupole moment, for
direct comparison with experiment.  One may observe that the present discussion represents a phenomenological rotational
analysis, in the traditional experimental sense, but of a large set of
observables taken from \textit{ab initio} calculations
of nuclei.  Having full access to the calculated wavefunctions, we may also
hope to extract information on the collective  structure
of the nuclear eigenstates from other measures of the
wave function correlations, such as density
distributions~\cite{cockrell2012:li-ncfc} and symmetry
decompositions~\cite{dytrych2008:sp-ncsm-deformation}.
Natural questions include the origin of rotation in the
$\isotope{Be}$ isotopes~--- for instance,
the extent to which it might arise from relative motion of alpha
clusters or perhaps from $\grpsu{3}$ rotation in an  extended multi-shell
valence space~--- and the relevance of some form of Nilsson-like
strong-coupling picture for rotation in \textit{ab initio}
calculations of odd-mass light nuclei.  Indeed, the proton and
neutron density distributions found in Ref.~\cite{maris2012:mfdn-ccp11}
for the ground state of $\isotope[9]{Be}$ suggest the emergence of two
alpha clusters, with the additional neutron in a $\pi$ orbital.


\section*{Acknowledgements} 

Discussions with A.~O.~Macchiavelli and P.~Fallon  are
gratefully acknowledged.
This work was supported by the Research Corporation for Science
Advancement through the Cottrell Scholar program, by the US Department
of Energy under Grants No.~DE-FG02-95ER-40934, DE-FC02-09ER41582
(SciDAC/UNEDF), DESC0008485
(SciDAC/NUCLEI), and DE-FG02-87ER40371, and by the US National Science
Foundation under Grant No.~0904782. Computational resources were
provided by the National Energy Research Supercomputer Center (NERSC),
which is supported by the Office of Science of the U.S. Department of
Energy under Contract No.~DE-AC02-05CH11231.


\section*{References}

\providecommand{\APSLONG}{}
\providecommand{\ELSEVIER}{}
\ELSEVIER



\clearpage



\begin{thebibliography}{41}
\expandafter\ifx\csname natexlab\endcsname\relax\def\natexlab#1{#1}\fi
\expandafter\ifx\csname bibnamefont\endcsname\relax
  \def\bibnamefont#1{#1}\fi
\expandafter\ifx\csname bibfnamefont\endcsname\relax
  \def\bibfnamefont#1{#1}\fi
\expandafter\ifx\csname citenamefont\endcsname\relax
  \def\citenamefont#1{#1}\fi
\expandafter\ifx\csname url\endcsname\relax
  \def\url#1{\texttt{#1}}\fi
\expandafter\ifx\csname urlprefix\endcsname\relax\def\urlprefix{URL }\fi
\providecommand{\bibinfo}[2]{#2}
\providecommand{\eprint}[2][arXiv]{\url{#1:#2}}
\renewcommand{\eprint}[2][arXiv]{\url{#1:#2}}

\bibitem{rowe2010:collective-motion}
\bibinfo{author}{\bibfnamefont{D.~J.} \bibnamefont{Rowe}},
  \emph{\bibinfo{title}{Nuclear Collective Motion: Models and Theory}}
  (\bibinfo{publisher}{World Scientific}, \bibinfo{address}{Singapore},
  \bibinfo{year}{2010}).

\bibitem{bohr1998:v1}
\bibinfo{author}{\bibfnamefont{A.}~\bibnamefont{Bohr}} \bibnamefont{and}
  \bibinfo{author}{\bibfnamefont{B.~R.} \bibnamefont{Mottelson}},
  \emph{\bibinfo{title}{Nuclear Structure}}, Vol.~\bibinfo{volume}{1}
  (\bibinfo{publisher}{World Scientific}, \bibinfo{address}{Singapore},
  \bibinfo{year}{1998}).

\bibitem{bohr1998:v2}
\bibinfo{author}{\bibfnamefont{A.}~\bibnamefont{Bohr}} \bibnamefont{and}
  \bibinfo{author}{\bibfnamefont{B.~R.} \bibnamefont{Mottelson}},
  \emph{\bibinfo{title}{Nuclear Structure}}, Vol.~\bibinfo{volume}{2}
  (\bibinfo{publisher}{World Scientific}, \bibinfo{address}{Singapore},
  \bibinfo{year}{1998}).

\bibitem{iachello1987:ibm}
\bibinfo{author}{\bibfnamefont{F.}~\bibnamefont{Iachello}} \bibnamefont{and}
  \bibinfo{author}{\bibfnamefont{A.}~\bibnamefont{Arima}},
  \emph{\bibinfo{title}{The Interacting Boson Model}}
  (\bibinfo{publisher}{Cambridge University Press},
  \bibinfo{address}{Cambridge}, \bibinfo{year}{1987}).

\bibitem{eisenberg1987:v1}
\bibinfo{author}{\bibfnamefont{J.~M.} \bibnamefont{Eisenberg}}
  \bibnamefont{and} \bibinfo{author}{\bibfnamefont{W.}~\bibnamefont{Greiner}},
  \emph{\bibinfo{title}{Nuclear Theory}}, \bibinfo{edition}{3rd} ed.,
  Vol.~\bibinfo{volume}{1} (\bibinfo{publisher}{North-Holland},
  \bibinfo{address}{Amsterdam}, \bibinfo{year}{1987}).

\bibitem{elliott1958:su3-part1}
\bibinfo{author}{\bibfnamefont{J.~P.} \bibnamefont{Elliott}},
  \bibinfo{journal}{Proc. R. Soc. London A} 245 (1958) 128.

\bibitem{harvey1968:su3-shell}
\bibinfo{author}{\bibfnamefont{M.}~\bibnamefont{Harvey}},
  \bibinfo{journal}{Adv. Nucl. Phys.} 1 (1968) 67.

\bibitem{pieper2004:gfmc-a6-8}
\bibinfo{author}{\bibfnamefont{S.~C.} \bibnamefont{Pieper}},
  \bibinfo{author}{\bibfnamefont{R.~B.} \bibnamefont{Wiringa}},
  \bibnamefont{and} \bibinfo{author}{\bibfnamefont{J.}~\bibnamefont{Carlson}},
  \bibinfo{journal}{Phys. Rev. C} 70 (2004) 054325.

\bibitem{neff2004:cluster-fmd}
\bibinfo{author}{\bibfnamefont{T.}~\bibnamefont{Neff}} \bibnamefont{and}
  \bibinfo{author}{\bibfnamefont{H.}~\bibnamefont{Feldmeier}},
  \bibinfo{journal}{Nucl. Phys. A} 738 (2004) 357.

\bibitem{hagen2007:coupled-cluster-benchmark}
\bibinfo{author}{\bibfnamefont{G.}~\bibnamefont{Hagen}},
  \bibinfo{author}{\bibfnamefont{D.~J.} \bibnamefont{Dean}},
  \bibinfo{author}{\bibfnamefont{M.}~\bibnamefont{Hjorth-Jensen}},
  \bibinfo{author}{\bibfnamefont{T.}~\bibnamefont{Papenbrock}},
  \bibnamefont{and} \bibinfo{author}{\bibfnamefont{A.}~\bibnamefont{Schwenk}},
  \bibinfo{journal}{Phys. Rev. C} 76 (2007) 044305.

\bibitem{navratil2009:ncsm}
\bibinfo{author}{\bibfnamefont{P.}~\bibnamefont{Navr\'{a}til}},
  \bibinfo{author}{\bibfnamefont{S.}~\bibnamefont{Quaglioni}},
  \bibinfo{author}{\bibfnamefont{I.}~\bibnamefont{Stetcu}}, \bibnamefont{and}
  \bibinfo{author}{\bibfnamefont{B.~R.} \bibnamefont{Barrett}},
  \bibinfo{journal}{J. Phys. G} 36 (2009) 083101.

\bibitem{bacca2012:6he-hyperspherical}
\bibinfo{author}{\bibfnamefont{S.}~\bibnamefont{Bacca}},
  \bibinfo{author}{\bibfnamefont{N.}~\bibnamefont{Barnea}}, \bibnamefont{and}
  \bibinfo{author}{\bibfnamefont{A.}~\bibnamefont{Schwenk}},
  \bibinfo{journal}{Phys. Rev. C} 86 (2012) 034321.

\bibitem{navratil2000:12c-ab-initio}
\bibinfo{author}{\bibfnamefont{P.}~\bibnamefont{Navr\'{a}til}},
  \bibinfo{author}{\bibfnamefont{J.~P.} \bibnamefont{Vary}}, \bibnamefont{and}
  \bibinfo{author}{\bibfnamefont{B.~R.} \bibnamefont{Barrett}},
  \bibinfo{journal}{Phys. Rev. Lett.} 84 (2000) 5728.

\bibitem{navratil2000:12c-ncsm}
\bibinfo{author}{\bibfnamefont{P.}~\bibnamefont{Navr\'{a}til}},
  \bibinfo{author}{\bibfnamefont{J.~P.} \bibnamefont{Vary}}, \bibnamefont{and}
  \bibinfo{author}{\bibfnamefont{B.~R.} \bibnamefont{Barrett}},
  \bibinfo{journal}{Phys. Rev. C} 62 (2000) 054311.

\bibitem{vary2009:ncsm-mfdn-scidac09}
\bibinfo{author}{\bibfnamefont{J.~P.} \bibnamefont{Vary}},
  \bibinfo{author}{\bibfnamefont{P.}~\bibnamefont{Maris}},
  \bibinfo{author}{\bibfnamefont{E.}~\bibnamefont{Ng}},
  \bibinfo{author}{\bibfnamefont{C.}~\bibnamefont{Yang}}, \bibnamefont{and}
  \bibinfo{author}{\bibfnamefont{M.}~\bibnamefont{Sosonkina}},
  \bibinfo{journal}{J. Phys. Conf. Ser.} 180 (2009) 012083.

\bibitem{barrett:ncsm}
\bibinfo{author}{\bibfnamefont{B.~R.} \bibnamefont{Barrett}},
  \bibinfo{author}{\bibfnamefont{P.}~\bibnamefont{Navr\'atil}},
  \bibnamefont{and} \bibinfo{author}{\bibfnamefont{J.~P.} \bibnamefont{Vary}},
  \bibinfo{journal}{Prog. Part. Nucl. Phys.}  (in press).

\bibitem{abe2012:fci-mcsm-ncfc}
\bibinfo{author}{\bibfnamefont{T.}~\bibnamefont{Abe}},
  \bibinfo{author}{\bibfnamefont{P.}~\bibnamefont{Maris}},
  \bibinfo{author}{\bibfnamefont{T.}~\bibnamefont{Otsuka}},
  \bibinfo{author}{\bibfnamefont{N.}~\bibnamefont{Shimizu}},
  \bibinfo{author}{\bibfnamefont{Y.}~\bibnamefont{Utsuno}}, \bibnamefont{and}
  \bibinfo{author}{\bibfnamefont{J.~P.} \bibnamefont{Vary}},
  \bibinfo{journal}{Phys. Rev. C} 86 (2012) 054301.

\bibitem{maris2009:ncfc}
\bibinfo{author}{\bibfnamefont{P.}~\bibnamefont{Maris}},
  \bibinfo{author}{\bibfnamefont{J.~P.} \bibnamefont{Vary}}, \bibnamefont{and}
  \bibinfo{author}{\bibfnamefont{A.~M.} \bibnamefont{Shirokov}},
  \bibinfo{journal}{Phys. Rev. C} 79 (2009) 014308.

\bibitem{bogner2008:ncsm-converg-2N}
\bibinfo{author}{\bibfnamefont{S.~K.} \bibnamefont{Bogner}},
  \bibinfo{author}{\bibfnamefont{R.~J.} \bibnamefont{Furnstahl}},
  \bibinfo{author}{\bibfnamefont{P.}~\bibnamefont{Maris}},
  \bibinfo{author}{\bibfnamefont{R.~J.} \bibnamefont{Perry}},
  \bibinfo{author}{\bibfnamefont{A.}~\bibnamefont{Schwenk}}, \bibnamefont{and}
  \bibinfo{author}{\bibfnamefont{J.}~\bibnamefont{Vary}},
  \bibinfo{journal}{Nucl. Phys. A} 801 (2008) 21.

\bibitem{cockrell2012:li-ncfc}
\bibinfo{author}{\bibfnamefont{C.}~\bibnamefont{Cockrell}},
  \bibinfo{author}{\bibfnamefont{J.~P.} \bibnamefont{Vary}}, \bibnamefont{and}
  \bibinfo{author}{\bibfnamefont{P.}~\bibnamefont{Maris}},
  \bibinfo{journal}{Phys. Rev. C} 86 (2012) 034325.

\bibitem{wiringa2000:gfmc-a8}
\bibinfo{author}{\bibfnamefont{R.~B.} \bibnamefont{Wiringa}},
  \bibinfo{author}{\bibfnamefont{S.~C.} \bibnamefont{Pieper}},
  \bibinfo{author}{\bibfnamefont{J.}~\bibnamefont{Carlson}}, \bibnamefont{and}
  \bibinfo{author}{\bibfnamefont{V.~R.} \bibnamefont{Pandharipande}},
  \bibinfo{journal}{Phys. Rev. C} 62 (2000) 014001.

\bibitem{neff2008:clustering-nuclei}
\bibinfo{author}{\bibfnamefont{T.}~\bibnamefont{Neff}} \bibnamefont{and}
  \bibinfo{author}{\bibfnamefont{H.}~\bibnamefont{Feldmeier}},
  \bibinfo{journal}{Eur. Phys. J. Special Topics} 156 (2008) 69.

\bibitem{kanadaenyo2012:amd-cluster}
\bibinfo{author}{\bibfnamefont{Y.}~\bibnamefont{Kanada-En`yo}},
  \bibinfo{author}{\bibfnamefont{M.}~\bibnamefont{Kimura}}, \bibnamefont{and}
  \bibinfo{author}{\bibfnamefont{A.}~\bibnamefont{Ono}},
  \bibinfo{journal}{Prog. Exp. Theor. Phys.} 2012 (2012) 01A202.

\bibitem{shimizu2012:mcsm}
\bibinfo{author}{\bibfnamefont{N.}~\bibnamefont{Shimizu}},
  \bibinfo{author}{\bibfnamefont{T.}~\bibnamefont{Abe}},
  \bibinfo{author}{\bibfnamefont{Y.}~\bibnamefont{Tsunoda}},
  \bibinfo{author}{\bibfnamefont{Y.}~\bibnamefont{Utsuno}},
  \bibinfo{author}{\bibfnamefont{T.}~\bibnamefont{Yoshida}},
  \bibinfo{author}{\bibfnamefont{T.~M.~M.} \bibnamefont{Honma}},
  \bibnamefont{and} \bibinfo{author}{\bibfnamefont{T.}~\bibnamefont{Otsuka}},
  \bibinfo{journal}{Prog. Exp. Theor. Phys.} 2012 (2012) 01A205.

\bibitem{maris2012:mfdn-ccp11}
\bibinfo{author}{\bibfnamefont{P.}~\bibnamefont{Maris}}, \bibinfo{journal}{J.
  Phys. Conf. Ser.} 402 (2012) 012031.

\bibitem{shirokov2007:nn-jisp16}
\bibinfo{author}{\bibfnamefont{A.~M.} \bibnamefont{Shirokov}},
  \bibinfo{author}{\bibfnamefont{J.~P.} \bibnamefont{Vary}},
  \bibinfo{author}{\bibfnamefont{A.~I.} \bibnamefont{Mazur}}, \bibnamefont{and}
  \bibinfo{author}{\bibfnamefont{T.~A.} \bibnamefont{Weber}},
  \bibinfo{journal}{Phys. Lett. B} 644 (2007) 33.

\bibitem{maris2012:mfdn-hites12}
\bibinfo{author}{\bibfnamefont{P.}~\bibnamefont{Maris}},
  \bibinfo{author}{\bibfnamefont{H.~M.} \bibnamefont{Aktulga}},
  \bibinfo{author}{\bibfnamefont{M.~A.} \bibnamefont{Caprio}},
  \bibinfo{author}{\bibfnamefont{U.~V.} \bibnamefont{Catalyurek}},
  \bibinfo{author}{\bibfnamefont{E.}~\bibnamefont{Ng}},
  \bibinfo{author}{\bibfnamefont{D.}~\bibnamefont{Oryspayev}},
  \bibinfo{author}{\bibfnamefont{H.}~\bibnamefont{Potter}},
  \bibinfo{author}{\bibfnamefont{E.}~\bibnamefont{Saule}},
  \bibinfo{author}{\bibfnamefont{M.}~\bibnamefont{Sosonkina}},
  \bibinfo{author}{\bibfnamefont{J.~P.} \bibnamefont{Vary}},
  \bibinfo{author}{\bibfnamefont{C.}~\bibnamefont{Yang}}, \bibnamefont{and}
  \bibinfo{author}{\bibfnamefont{Z.}~\bibnamefont{Zhou}}, \bibinfo{journal}{J.
  Phys. Conf. Ser.} 403 (2012) 012019.

\bibitem{sternberg2008:ncsm-mfdn-sc08}
\bibinfo{author}{\bibfnamefont{P.}~\bibnamefont{Sternberg}},
  \bibinfo{author}{\bibfnamefont{E.~G.} \bibnamefont{Ng}},
  \bibinfo{author}{\bibfnamefont{C.}~\bibnamefont{Yang}},
  \bibinfo{author}{\bibfnamefont{P.}~\bibnamefont{Maris}},
  \bibinfo{author}{\bibfnamefont{J.~P.} \bibnamefont{Vary}},
  \bibinfo{author}{\bibfnamefont{M.}~\bibnamefont{Sosonkina}},
  \bibnamefont{and} \bibinfo{author}{\bibfnamefont{H.~V.} \bibnamefont{Le}}, in
  \emph{\bibinfo{booktitle}{SC '08: Proceedings of the 2008 ACM/IEEE Conference
  on Supercomputing}} (\bibinfo{publisher}{IEEE Press},
  \bibinfo{address}{Piscataway, NJ}, \bibinfo{year}{2008}),
  \bibinfo{note}{{A}rticle No. 15}.

\bibitem{maris2010:ncsm-mfdn-iccs10}
\bibinfo{author}{\bibfnamefont{P.}~\bibnamefont{Maris}},
  \bibinfo{author}{\bibfnamefont{M.}~\bibnamefont{Sosonkina}},
  \bibinfo{author}{\bibfnamefont{J.~P.} \bibnamefont{Vary}},
  \bibinfo{author}{\bibfnamefont{E.}~\bibnamefont{Ng}}, \bibnamefont{and}
  \bibinfo{author}{\bibfnamefont{C.}~\bibnamefont{Yang}},
  \bibinfo{journal}{Procedia Comput. Sci.} 1 (2010) 97.

\bibitem{aktulga2012:mfdn-ep2012}
\bibinfo{author}{\bibfnamefont{H.~M.} \bibnamefont{Aktulga}},
  \bibinfo{author}{\bibfnamefont{C.}~\bibnamefont{Yang}},
  \bibinfo{author}{\bibfnamefont{E.~G.} \bibnamefont{Ng}},
  \bibinfo{author}{\bibfnamefont{P.}~\bibnamefont{Maris}}, \bibnamefont{and}
  \bibinfo{author}{\bibfnamefont{J.~P.} \bibnamefont{Vary}}, in
  \emph{\bibinfo{booktitle}{Euro-Par 2012 Parallel Processing}}, edited by
  \bibinfo{editor}{\bibfnamefont{C.}~\bibnamefont{Kaklamanis}},
  \bibinfo{editor}{\bibfnamefont{T.}~\bibnamefont{Papatheodorou}},
  \bibnamefont{and} \bibinfo{editor}{\bibfnamefont{P.~G.}
  \bibnamefont{Spirakis}} (\bibinfo{publisher}{Springer-Verlag},
  \bibinfo{address}{Berlin}, \bibinfo{year}{2012}), \bibinfo{series}{Lecture
  Notes in Computer Science} Vol. \bibinfo{volume}{7484}, p.
  \bibinfo{pages}{830}.

\bibitem{npa1990:011-012}
\bibinfo{author}{\bibfnamefont{F.}~\bibnamefont{Ajzenberg-Selove}},
  \bibinfo{journal}{Nucl. Phys. A} 506 (1990) 1.

\bibitem{npa2002:005-007}
\bibinfo{author}{\bibfnamefont{D.~R.} \bibnamefont{Tilley}},
  \bibinfo{author}{\bibfnamefont{C.~M.} \bibnamefont{Cheves}},
  \bibinfo{author}{\bibfnamefont{J.~L.} \bibnamefont{Godwin}},
  \bibinfo{author}{\bibfnamefont{G.~M.} \bibnamefont{Hale}},
  \bibinfo{author}{\bibfnamefont{H.~M.} \bibnamefont{Hofmann}},
  \bibinfo{author}{\bibfnamefont{J.~H.} \bibnamefont{Kelley}},
  \bibinfo{author}{\bibfnamefont{C.~G.} \bibnamefont{Sheu}}, \bibnamefont{and}
  \bibinfo{author}{\bibfnamefont{H.~R.} \bibnamefont{Weller}},
  \bibinfo{journal}{Nucl. Phys. A} 708 (2002) 3.

\bibitem{nilsson1955:model}
\bibinfo{author}{\bibfnamefont{S.~G.} \bibnamefont{Nilsson}},
  \bibinfo{journal}{Mat. Fys. Medd. Dan. Vid. Selsk.} 29~(16) (1955).

\bibitem{mottelson1959:nilsson-model}
\bibinfo{author}{\bibfnamefont{B.~R.} \bibnamefont{Mottelson}}
  \bibnamefont{and} \bibinfo{author}{\bibfnamefont{S.~G.}
  \bibnamefont{Nilsson}}, \bibinfo{journal}{Mat. Fys. Skr. Dan. Vid. Selsk.}
  1~(8) (1959).

\bibitem{weisskopf1951:estimate}
\bibinfo{author}{\bibfnamefont{V.~F.} \bibnamefont{Weisskopf}},
  \bibinfo{journal}{Phys. Rev.} 83 (1951) 1073.

\bibitem{draayer1984:spsm-20ne}
\bibinfo{author}{\bibfnamefont{J.~P.} \bibnamefont{Draayer}},
  \bibinfo{author}{\bibfnamefont{K.~J.} \bibnamefont{Weeks}}, \bibnamefont{and}
  \bibinfo{author}{\bibfnamefont{G.}~\bibnamefont{Rosensteel}},
  \bibinfo{journal}{Nucl. Phys. A} 413 (1984) 215.

\bibitem{rowe1985:micro-collective-sp6r}
\bibinfo{author}{\bibfnamefont{D.~J.} \bibnamefont{Rowe}},
  \bibinfo{journal}{Rep. Prog. Phys.} 48 (1985) 1419.

\bibitem{dytrych2008:sp-ncsm}
\bibinfo{author}{\bibfnamefont{T.}~\bibnamefont{Dytrych}},
  \bibinfo{author}{\bibfnamefont{K.~D.} \bibnamefont{Sviratcheva}},
  \bibinfo{author}{\bibfnamefont{J.~P.} \bibnamefont{Draayer}},
  \bibinfo{author}{\bibfnamefont{C.}~\bibnamefont{Bahri}}, \bibnamefont{and}
  \bibinfo{author}{\bibfnamefont{J.~P.} \bibnamefont{Vary}},
  \bibinfo{journal}{J. Phys. G} 35 (2008) 123101.

\bibitem{furnstahl2012:ho-extrapolation}
\bibinfo{author}{\bibfnamefont{R.~J.} \bibnamefont{Furnstahl}},
  \bibinfo{author}{\bibfnamefont{G.}~\bibnamefont{Hagen}}, \bibnamefont{and}
  \bibinfo{author}{\bibfnamefont{T.}~\bibnamefont{Papenbrock}},
  \bibinfo{journal}{Phys. Rev. C} 86 (2012) 031301.

\bibitem{coon2012:nscm-ho-regulator}
\bibinfo{author}{\bibfnamefont{S.~A.} \bibnamefont{Coon}},
  \bibinfo{author}{\bibfnamefont{M.~I.} \bibnamefont{Avetian}},
  \bibinfo{author}{\bibfnamefont{M.~K.~G.} \bibnamefont{Kruse}},
  \bibinfo{author}{\bibfnamefont{U.}~\bibnamefont{van Kolck}},
  \bibinfo{author}{\bibfnamefont{P.}~\bibnamefont{Maris}}, \bibnamefont{and}
  \bibinfo{author}{\bibfnamefont{J.~P.} \bibnamefont{Vary}},
  \bibinfo{journal}{Phys. Rev. C} 86 (2012) 054002.

\bibitem{dytrych2008:sp-ncsm-deformation}
\bibinfo{author}{\bibfnamefont{T.}~\bibnamefont{Dytrych}},
  \bibinfo{author}{\bibfnamefont{K.~D.} \bibnamefont{Sviratcheva}},
  \bibinfo{author}{\bibfnamefont{C.}~\bibnamefont{Bahri}},
  \bibinfo{author}{\bibfnamefont{J.~P.} \bibnamefont{Draayer}},
  \bibnamefont{and} \bibinfo{author}{\bibfnamefont{J.~P.} \bibnamefont{Vary}},
  \bibinfo{journal}{J. Phys. G} 35 (2008) 095101.

\end{thebibliography}
\end{document}